\documentclass{an-think}

\usepackage{graphicx,fancyhdr}
\usepackage{times}

\begin{document}

\sloppy
\newcommand{\kms}{km\,s$^{-1}$}
\newcommand{\Halpha}{H$\alpha$}


\title{     SDSS Data Management and Photometric Quality Assessment  }

\author{
\v{Z} Ivezi\'{c}\inst{1,}\inst{2}, 
R.H. Lupton\inst{1}
\and
D. Schlegel\inst{1}
\and
B. Boroski\inst{3}
\and
J. Adelman-McCarthy\inst{3}
\and 
B. Yanny\inst{3}
\and 
S. Kent\inst{3}
\and 
C. Stoughton\inst{3}
\and
D. Finkbeiner\inst{1}
\and
N. Padmanabhan\inst{4}
\and 
C.M. Rockosi\inst{5}
\and
J.E. Gunn\inst{1}
\and
G.R. Knapp\inst{1}
\and 
M.A. Strauss\inst{1}
\and
G.T. Richards\inst{1}
\and
D. Eisenstein\inst{6}
\and
T. Nicinski\inst{7}
\and
S.J. Kleinman\inst{8}
\and
J. Krzesinski\inst{8}
\and
P.R. Newman\inst{8}
\and
S. Snedden\inst{8}
\and
A.R. Thakar\inst{9}
\and
A. Szalay\inst{9}
\and
J.A. Munn\inst{10}
\and
J.A. Smith\inst{11}
\and 
D. Tucker\inst{3}
\and
B.C. Lee\inst{12}
\\}

\institute{
Princeton University Observatory, Princeton, NJ 08544
\and
H.N. Russell Fellow, on leave from the University of Washington (e-mail:
ivezic@astro.princeton.edu)
\and
Fermi National Accelerator Laboratory, P.O. Box 500, Batavia, IL 60510
\and
Princeton University, Dept. of Physics, Princeton, NJ 08544
\and
University of Washington, Dept. of Astronomy, Box 351580, Seattle, WA 98195
\and
Steward Observatory, 933 N. Cherry Ave., Tucson, AZ 85721
\and
CMC Electronics Aurora, 43W752 Route 30, Sugar Grove, IL 60554
\and
Apache Point Observatory, 2001 Apache Point Road, P.O. Box 59, Sunspot, NM 88349-0059
\and
Department of Physics and
Astronomy, The John Hopkins University, 3701 San Martin Drive, Baltimore, MD 21218
\and
U.S. Naval Observatory, Flagstaff Station, P.O. Box 1149, Flagstaff, AZ 86002
\and
Space Instr. \& Systems Engineering, ISR-4, MS D448, Los Alamos National Laboratory
Los Alamos, NM 87545
\and
Lawrence Berkeley National Laboratory, One Cyclotron Road, MS 50R5032, Berkeley, CA, 94720 
}

\date{Received; accepted; published online}


\abstract{ 
We summarize the Sloan Digital Sky Survey data acquisition and processing 
steps, and describe
{\it runQA}, a pipeline designed for automated data quality assessment. In 
particular, we show how the position of the stellar locus in color-color
diagrams can be used to estimate the accuracy of photometric zeropoint 
calibration to better than 0.01 mag in 0.03 deg$^2$ patches. Using
this method, we estimate that typical photometric zeropoint calibration 
errors for SDSS imaging data are not larger than $\sim0.01$ mag in the 
$g$, $r$, and $i$ bands, 0.02 mag in the $z$ band, and 0.03 mag in the 
$u$ band (root-mean-scatter for zeropoint offsets). 
}

\maketitle


\section{Introduction}

Modern large-scale digital sky surveys, such as SDSS, 2MASS, and FIRST, 
are opening new frontiers in astronomy. Due to large 
data rates, they require sophisticated data acquisition, processing
and distribution systems. The quality of data is of paramount importance
for their scientific impact, but its quantitative assessment is a difficult
problem -- mainly because it is not known very precisely what measurement 
values to expect. In the majority of large surveys to date, data quality
assessment has not been fully automated and has required substantial human intervention
-- a significant shortcoming and resource sink when the data rate and volume are large. 

The SDSS (Sloan Digital Sky Survey, York et al.~2000, Abazajian et al.~2003) 
is an optical imaging and spectroscopic survey, which aims to 
cover one quarter of the sky, and obtain high-quality spectra for
100,000 quasars, a similar number of stars, and a million galaxies.
The SDSS observing, data processing, and data dissemination are highly
automated operations -- occasionally, SDSS is referred to as a {\it science
factory} (to date, over a thousand journal papers are based on, or refer
to, SDSS). We provide a brief summary of these operations in Section 2. 

As has been the case for other surveys with large data rates, the data quality assessment 
and assurance has been a hard problem for SDSS. Recently, we have 
developed a set of tools, organized in the {\it runQA} pipeline, that are
used to automatically assess the quality of SDSS imaging data products, and
report all instances where it is substandard. The main ideas and results
are described in Section 3, and further discussed in Section 4.

\section{     SDSS data flow and processing                        }

\subsection{Overview of SDSS imaging data}

SDSS is providing homogeneous and deep ($r < 22.5$) photometry in five pass-bands 
($u$, $g$, $r$, $i$, and $z$, Fukugita et al.~1996; Gunn et al.~1998; Smith 
et al.~2002; Hogg et al. 2002) accurate to 0.02 mag (rms, for sources not limited
by photon statistics, Ivezi\'{c} et al.~2003). 
The survey sky coverage of 10,000 deg$^2$ in the Northern Galactic Cap,
and $\sim 200$ deg$^2$ in the Southern Galactic Hemisphere, will result in photometric 
measurements for over 100 million stars and a similar number of galaxies. 
Astrometric positions are accurate to better than 0.1 arcsec per coordinate (rms) 
for sources with $r<20.5^m$ (Pier et al.~2003), and the morphological information 
from the images allows reliable star-galaxy separation to $r \sim$ 21.5$^m$ (Lupton 
et al.~2002).

\subsection{Data acquisition}

The data acquisition system (Petravick et al. 1994) records information from the 
imaging camera, spectrographs, and photometric telescope (used to obtain photometric 
calibration data). The imaging camera produces the highest data rate ($\sim$ 20 GB/hour). 
Each system uses report files to track the observations.

Data from the imaging camera (thirty photometric, twelve astrometric, and two focus CCDs, 
Gunn et al. 1998)
are collected in the drift scan mode. The images that correspond to the same sky location in each of 
the five photometric filters (these five images are collected over $\sim$5 minutes, with 54 sec 
per individual
exposure) are grouped together for processing as a field. Frames from the astrometric and focus
CCDs are not saved, but rather, stars from them are detected and measured in real time to provide 
feedback on telescope tracking and focus. This same 
analysis is done for the photometric CCDs, and we save these results along with the actual frames. 

Data from the spectrographs are read from the four CCDs (one red channel and one blue channel in each
of the two spectrographs) after each exposure. A complete set of exposures includes bias, flat, arc, 
and science exposures taken through the fibers, as well as a uniformly illuminated flat to
take out pixel-to-pixel variations.

Data from the photometric telescope (PT) include bias frames, dome and twilight flats 
for each filter, measurements of primary standards in each filter, and measurements of 
secondary calibration patches (calibrated using primary standards, and sufficiently 
faint to be unsaturated in the main survey data) in each filter.

All of these systems are supported by a common set of observers' programs, with observer 
interfaces customized for each system to optimize the observing efficiency.

\subsection{ Data processing factory }

Data from Apache Point Observatory (APO) are transferred to Fermilab for processing and calibration
via magnetic tape (using a commercial carrier), with critical, low-volume samples sent over 
the Internet. Imaging data are processed with the imaging pipelines: 
the astrometric pipeline ({\it astrom}) performs the astrometric calibration (Pier et al. 2003); 
the postage-stamp 
pipeline ({\it psp}) characterizes the behavior of the point-spread function (PSF) as a function
of time and location 
in the focal plane; the frames pipeline ({\it frames}) finds, deblends, and measures the properties 
of objects; and the final calibration pipeline ({\it nfcalib}) applies the photometric calibration 
to the objects. This calibration uses the results of the PT data processed with the monitor telescope
pipeline ({\it mtpipe}). The combination of the psp and frames pipelines is sometimes referred to as 
{\it photo} (Lupton et al. 2002). 

Individual imaging runs that interleave are prepared for spectroscopy with the following steps: 
{\it resolve} selects a primary detection for objects that fall in an overlap area; the target 
selection pipeline ({\it target}) selects objects for spectroscopic observation 
(for galaxies see Strauss et al. 2002, for quasars Richards et al. 2002, and for luminous
red galaxies Eisenstein et al. 2002); and the plate 
pipeline ({\it plate}) specifies the locations of the plates on the sky and the location of holes 
to be drilled in each plate (Blanton et al. 2003). Spectroscopic data are first extracted and 
calibrated with the two-dimensional pipeline ({\it spectro2d}) and then classified and
measured with the one-dimensional pipeline ({\it spectro1d}).

A compendium of technical details about individual pipelines and data processing can be found
in Stoughton et al. (2002), Abazajian et al. (2003, 2004), and on the SDSS web site (www.sdss.org).

\subsection{              Data dissemination                  }

There are three database servers that can be used to access the public imaging and spectroscopic 
SDSS data. The searchable Catalog Archive Server and Skyserver contain the measured and calibrated
parameters from all
objects in the imaging survey and the spectroscopic survey. The Data Archive Server contains 
the rest of the data products, such as the corrected imaging frames
and the calibrated spectra. More details about user interfaces  can be found 
on the SDSS web site (www.sdss.org).

\section{                   Photometric quality assurance             }

SDSS imaging data are photometrically calibrated using a network
of calibration stars obtained in $\sim$2 degree large patches by 
the Photometric Telescope (Smith et al.~2002). The quality of SDSS photometry stands 
out among available large-area optical sky surveys 
(Ivezi\'{c} et al.~2003, Sesar et al.~2004). Nevertheless, the achieved accuracy
is occasionally worse than the nominal 0.02 mag (root-mean-square for sources not limited 
by photon statistics). Typical causes of substandard photometry include an incorrectly
modeled PSF (usually due to fast changing atmospheric seeing, 
or lack of a sufficient number of the isolated bright stars needed for modeling), unrecognized
changes in atmospheric transparency, errors in photometric zeropoint calibration,
effects of crowded fields at low Galactic latitudes, undersampled PSF in excellent seeing 
conditions ($\la 0.8$ arcsec; the pixel size is 0.4 arcsec), incorrect flatfield, or 
bias vectors, scattered light etc. 
Such effects can conspire to increase the photometric errors to levels as high as 0.05 mag. 

It is desirable to recognize and record all instances of substandard photometry, 
as well as to routinely track the overall data integrity. Due to the high data volume,
such procedures need to be fully automated. In this Section we describe such 
procedures, implemented in the {\it runQA} pipeline, for tracking the accuracy of PSF photometry 
and photometric zeropoint calibration.

\subsection{ Point spread function photometry }

The PSF flux is computed using the PSF as a weighting function.
While this flux is optimal for faint point sources (in particular,
it is vastly superior to aperture photometry at the faint end), it is also
sensitive to inaccurate PSF modeling, which attempts to capture the complex
PSF behavior. Even in the absence of atmospheric inhomogeneities, 
the SDSS telescope delivers images whose FWHMs vary by up to 15\% from one side 
of a CCD to the other; the worst effects are seen in the chips farthest from the 
optical axis. Moreover, since the atmospheric seeing varies with time, the delivered 
image quality is a complex two-dimensional function even on the scale of a single frame. 
Without an accurate model, the PSF photometry would have errors up to 0.10-0.15 mag.
The description of the point-spread function is also critical for star-galaxy separation
and for unbiased measures of the shapes of nonstellar objects.

The SDSS imaging PSF is modeled heuristically in each band using a Karhunen-Loeve 
(K-L) transform (Lupton et al. 2002). Using stars brighter than 
roughly 20$^{\rm th}$ magnitude, the PSF from a series of five frames is expanded into 
eigenimages and the first three terms are retained. The variation of these coefficients
is then fit up to a second order polynomial in each chip coordinate.

The success of this K-L expansion (a part of the {\it psp} pipeline) is gauged by 
comparing PSF photometry based on the modeled K-L PSFs with large-aperture 
photometry for the same (bright) stars. In addition to initial comparison
implemented in {\it psp}, the quality assurance pipeline {\it runQA} performs
the same analysis using outputs from the {\it frames} pipeline, which have 
more accurately measured object parameters and more robust star/galaxy separation.

\begin{figure}[htb]
   \centering
   \includegraphics[angle=0,width=8cm]{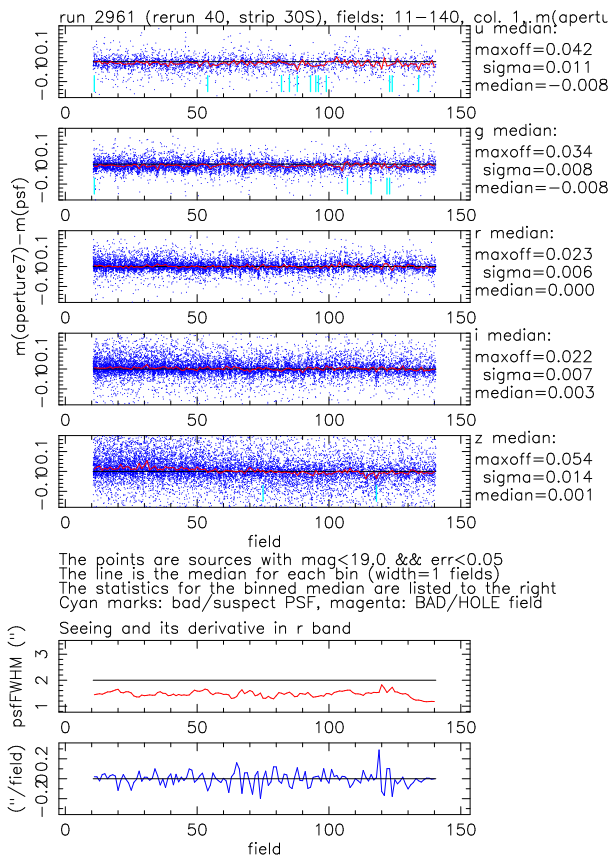}
   \caption{An example of output from the SDSS photometric assurance pipeline
 {\it runQA}. The top five panels show the difference between aperture and
  PSF magnitudes of stars brighter than $m\sim$19, as a function of time 
  (1 field = 36 sec), for five SDSS bands. 
  The statistics for medians, evaluated for each field, are shown next 
  to each panel ({\it maxoff} is the maximum deviation from zero, {\it sigma} is
  the rms scatter). The vertical lines in each panel mark bad fields.
  The bottom two panels show the image quality in the
  $r$ band (FWHM and its time derivative) as a function of time.} 
   \label{aperPSF}
\end{figure}

Typical behavior of the difference between aperture and PSF magnitudes
as a function of time is shown in Fig.~\ref{aperPSF}. The low-order statistics
of medians, evaluated for each field and band, are used to recognize and flag
all fields with substandard PSF photometry. Such ``bad'' fields usually also
have substandard star-galaxy separation and other measurements that critically
depend on the accurate PSF model.

\subsection{ Photometric zeropoint calibration}

Data from each camera column, and for a given run, are independently calibrated. 
The calibration pipeline {\it nfcalib}
reports the rms (dis)agreement, which is typically $\sim$0.02 mag
(the core width, but the distribution is not necessarily Gaussian). There 
are usually several calibration patches for a given run, which are separated
by of order an hour of scanning time. Thus, any changes in 
atmospheric transparency, or other conditions affecting the photometric
sensitivity, may not be recognized on shorter timescales. In addition,
the stability of photometric calibration across the sky is sensitive
to systematic errors in the calibration star network. Hence, it is 
desirable to have an independent estimate of the stability of 
photometric calibration across the sky, as well as of its 
behavior on short time scales (say, a few minutes).

A dense network of calibration stars ($\ga$100 stars per deg$^{2}$) across
the sky, accurate to $\sim$0.01 mag, in five SDSS bands, which could
be used for an independent verification of SDSS photometric calibration,
does not yet exist. Fortunately, the distribution of stars in SDSS
color-color diagrams seems fairly stable across the sky ($<$0.01 mag),
and offers an indirect but powerful test of photometric zeropoint calibration.

\subsubsection{ Definition of the stellar locus in the SDSS photometric system  }

The majority of stars detected by SDSS are on the main sequence ($>$98\%, Finlator 
et al. 2001, Helmi et al. 2002). They form a well defined sequence, usually referred
to as a ``stellar locus'', in color-color diagrams (Lenz et al, 1998, Fan 1999, 
Smol\v{c}i\'{c} et al. 2004). The particular morphology displayed by the stellar 
locus can be used to measure whether it is ``in the same place'' for independently
calibrated data.

The methodology used to derive the principal colors which track the position
of the stellar locus is described in Helmi et al. (2002). Briefly, two principal
axes, $P_1$ and $P_2$, are defined along the locus and perpendicular to the locus, 
for the appropriately chosen parts of the locus, and in three color-color planes 
spanned by SDSS photometric system. The color perpendicular to the locus, $P_2$, is 
adjusted for a small dependence on apparent magnitude ($<$0.01 mag/mag), to obtain
$P'_2$, which is then used for high-precision tracking of
the locus position.

\begin{figure}[t]
   \centering
   \includegraphics[angle=0,width=8cm]{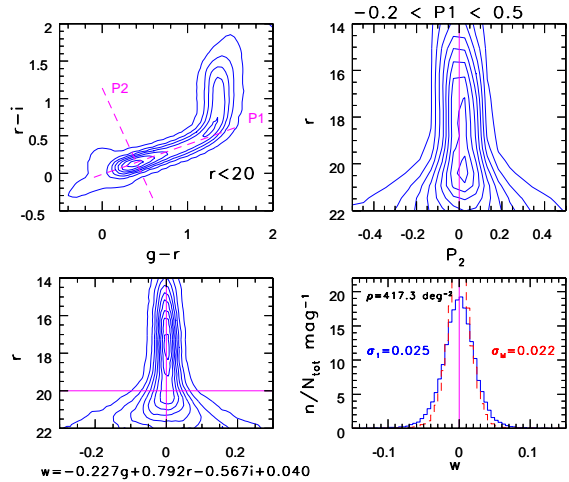}
   \caption{An example of the definition of principal color axes in 
   color-color diagrams. The top left panel shows the $r-i$ vs. 
   $g-r$ color-color diagram, with the $P_1$ axis along the blue part 
   of the locus, and $P_2$ perpendicular to the locus. The top right
   panel shows the $P_2$ color as a function of $r$ magnitude. The 
   $w$ principal color, shown as a function of $r$ magnitude in the 
   bottom left panel, is obtained by correcting $P_2$ for its small 
   dependence on $r$, and renormalizing it such that $w$ error (random,
   not systematic) is
   comparable to the mean error in the $g$, $r$, and $i$ bands. The solid 
   histogram in the bottom right panel shows the distribution of $w$ color 
   for stars with $r<20$. The dashed histogram shows the distribution
   of $w$ color constructed with the mean of five  measurements (i.e.
   five passes over a given region of sky). 
   The distribution rms width decreases from 0.025 mag to 0.022 mag, 
   which implies that single measurement error for the $w$ color is
   $\sim$0.01 mag, and that the intrinsic locus width in the $w$ 
   direction is $\sim$0.02 mag.} 
   \label{wcolordef}
\end{figure}

There are four principal colors used by {\it runQA}: $s$ (the blue part of 
the locus in the $g$-$r$ vs. $u$-$g$ plane), $w$ (blue part in $r$-$i$ vs. 
$g$-$r$), $x$ (red part in $r$-$i$ vs. $g$-$r$), and $y$ (red part in $i$-$z$ vs. $r$-$i$).  
As an example, the definition and properties of the $w$ color are shown in 
Fig.~\ref{wcolordef}. The principal colors are defined by the following linear 
combinations of magnitudes in the five SDSS bands:
\begin{equation}
          P'_2 = A\,u + B\,g + C\,r + D\,i + E\,z + F,
\end{equation}
and
\begin{equation}
           P_1 = A'\,u + B'\,g + C'\,r + D'\,i + E'\,z + F',
\end{equation}
where $P'_2 = s, w, x, y$ are the principal colors perpendicular to the
locus in a particular color-color diagram, and $P_1$ (for each $P'_2$) are 
the corresponding principal axes along the locus. All the measurements
are corrected for interstellar extinction using the Schlegel, Finkbeiner
\& Davis (1998) map (SFD). The adopted coefficients 
$A-F$ and $A'-F'$ are listed in Tables 1 and 2. Stars used for estimating the position 
of the stellar locus must not have processing flags BRIGHT, SATUR, and BLENDED 
set (see Stoughton et al. 2002 for more details about {\it photo} processing flags), 
and must also satisfy $r < r_{max}$ and $P_1^{min} < P_1 < P_1^{max}$, 
with $r_{max}$, $P_1^{min}$ and $P_1^{max}$ for each principal color listed in Table 3.
The typical width of the stellar locus (the rms distribution width for each 
principal color) is listed in the last column in Table 3. 

Even at high galactic latitudes ($|b|>30$), the surface density of stars 
satisfying these criteria is sufficient for evaluating the mean of the principal
colors with an error of $\sim0.01$ mag per SDSS field (area of 0.032 deg$^2$). 
The mean error values per SDSS field for each principal color are listed in 
Table 3 ($\sigma$). The principal colors are evaluated by {\it runQA} in four 
field wide bins, yielding errors twice as small as those listed in Table 3.

The achievable accuracy in the determination of the position of the 
stellar locus depends on the number of stars in the sample and the 
intrinsic locus width. While the measured locus width is broadened by 
photometric errors, its value is, nevertheless, dominated by the intrinsic
distribution of stellar properties such as metallicity and surface gravity. 
This conclusion is based on a comparison of the locus width for a set 
of single measurements, and using the mean values for five measurements of the
same stars (see the bottom right panel in Fig.~\ref{wcolordef}). Although
the latter data have much smaller measurement errors, the width of the 
distribution is not significantly decreased, showing that most of the 
observed width is intrinsic. 

\begin{table}
  \centering
   \caption{The $P'_2$=$s, w, x, y$ Principal Color Definitions}
   \begin{tabular}{@{}ccccccc@{}}
    \hline
   $P'_2$  &    A   &    B   &    C   &    D   &    E   &   F     \\
    	\hline
   s   & -0.249 &  0.794 & -0.555 &  0.0   &  0.0   &  0.234 \\ 
   w   &  0.0   & -0.227 &  0.792 & -0.567 &  0.0   &  0.050 \\
   x   &  0.0   &  0.707 & -0.707 &  0.0   &  0.0   & -0.988 \\
   y   &  0.0   &  0.0   & -0.270 &  0.800 & -0.534 &  0.054 \\
 	\hline
   \end{tabular}
 \end{table}

\begin{table}
  \centering
   \caption{The $P_1$ Principal Color Definitions for each $P'_2$}
   \begin{tabular}{@{}ccccccc@{}}
    \hline
   $P'_2$  &    A'  &    B'  &    C'  &    D'  &    E'  &   F'     \\
    	\hline
   s   &  0.910 &  -0.495 & -0.415 &  0.0    &  0.0   &  -1.28   \\ 
   w   &  0.0   &   0.928 & -0.556 & -0.372  &  0.0   &  -0.425   \\
   x   &  0.0   &  0.0    &  1.0   &  -1.0   &  0.0   &   0.0    \\
   y   &  0.0   &  0.0    &  0.895 &  -0.448 & -0.447 &  -0.600   \\
 	\hline
   \end{tabular}
 \end{table}

\begin{table}
  \centering
   \caption{Additional Constraints for the Principal Color Definitions}
   \begin{tabular}{@{}cccccc@{}}
    \hline
 $P'_2$ &  $r_{max}$ & $P_1^{min}$ & $P_1^{max}$  & $\sigma$  (mag) & width (mag) \\
    	\hline
 s &    19.0    &   -0.2        &   0.8           &  0.011   &  0.031  \\
 w &    20.0    &   -0.2        &   0.6           &  0.006   &  0.025  \\
 x &    19.0    &    0.8        &   1.6           &  0.021   &  0.042  \\
 y &    19.5    &    0.1        &   1.2           &  0.008   &  0.023  \\
	\hline
   \end{tabular}

 \end{table}

\subsubsection{ The measurements of the position of the stellar locus  }

For each of the four principal colors, the {\it runQA} pipeline computes
and reports the low-order distribution statistics (such as median, rms,
and the fraction of 3$\sigma$ outliers), in graphical and tabular
form. These statistics are used to evaluate the accuracy of photometric
zeropoint calibration for each run/camera column combination, as well
as to track photometric accuracy as a function of time. For example,
in case of unrecognized non-gray atmospheric extinction variations, the
position of the locus (the principal color medians) changes, too. 
When the PSF model is incorrect, the locus width (and typically the fraction 
of 3$\sigma$ outliers) increases. As an example, the behavior of the 
$w$ color as a function of time in one run is shown in Fig.~\ref{wcolortime}.
All instances where the median principal color, or principal color 
distribution width, are outside adopted boundaries are interpreted 
as fields with substandard photometry, and reported in a ``field
quality'' table. The distributions of the median principal colors 
for 291 SDSS runs processed to date are shown in the left two panels
in Fig.~\ref{PCdistribution}. The mean distribution widths are listed
in the last column in Table 3. 

In principle, the median principal colors could vary across the sky
due to galactic structure and stellar population effects. In order
to estimate the magnitude of such variations, we consider the 
rms scatter of the mean stellar locus position for six
camera columns in a given run, and its distribution for all runs.
This scatter should be insensitive to the exact position of the
stellar locus because stellar population variations should not be 
significant on the camera's angular scale of $\sim$2 degree. 
If the position of the stellar locus does not vary appreciably
across the sky, then the median value of this rms scatter should
be similar to the rms scatter of the median position of the stellar
locus. We find this to be the case for all four principal colors
(see the right two panels in Fig.~\ref{PCdistribution}).
This agreement suggests that the variations in the median principal 
colors are dominated by the errors in photometric zeropoint calibration,
although we can not exclude the possibility that the observed variations
are at least partly due to galactic structure effects. Hence,
the estimates of photometric zeropoint calibration errors presented
here are upper limits, and the true calibration errors could be
indeed somewhat smaller.

While SDSS is by design a high galactic latitude survey, there are
imaging data that probe low galactic latitudes. We find that the
method for tracking the position of the stellar locus described here 
starts to fail at latitudes below $|b|=10-15$ deg. The main reason 
is that nearby M dwarfs are not behind all the dust implied by 
the SFD extinction map. As the extinction increases towards the
Galactic plane, this fact becomes increasingly important, and 
induces a shift in the position of the stellar locus. While it 
may be possible to account for this effect by adopting a photometric
parallax relation for M dwarfs, and assuming a dust distribution model, 
this possibility has not yet been quantitatively investigated.

\begin{figure}[t]
   \centering
   \includegraphics[angle=0,width=8cm]{ivezic.fig3.ps}
   \caption{An example of the stellar locus position (top), rms width 
   (second panel from top), and tail behavior (the bottom two panels)
   as a function of time, for the six camera columns. The ``tails''
   are defined as 2$\sigma$ outliers.} 
   \label{wcolortime}
\end{figure}

\begin{figure*}[t]
   \centering
   \includegraphics[angle=0,width=16cm]{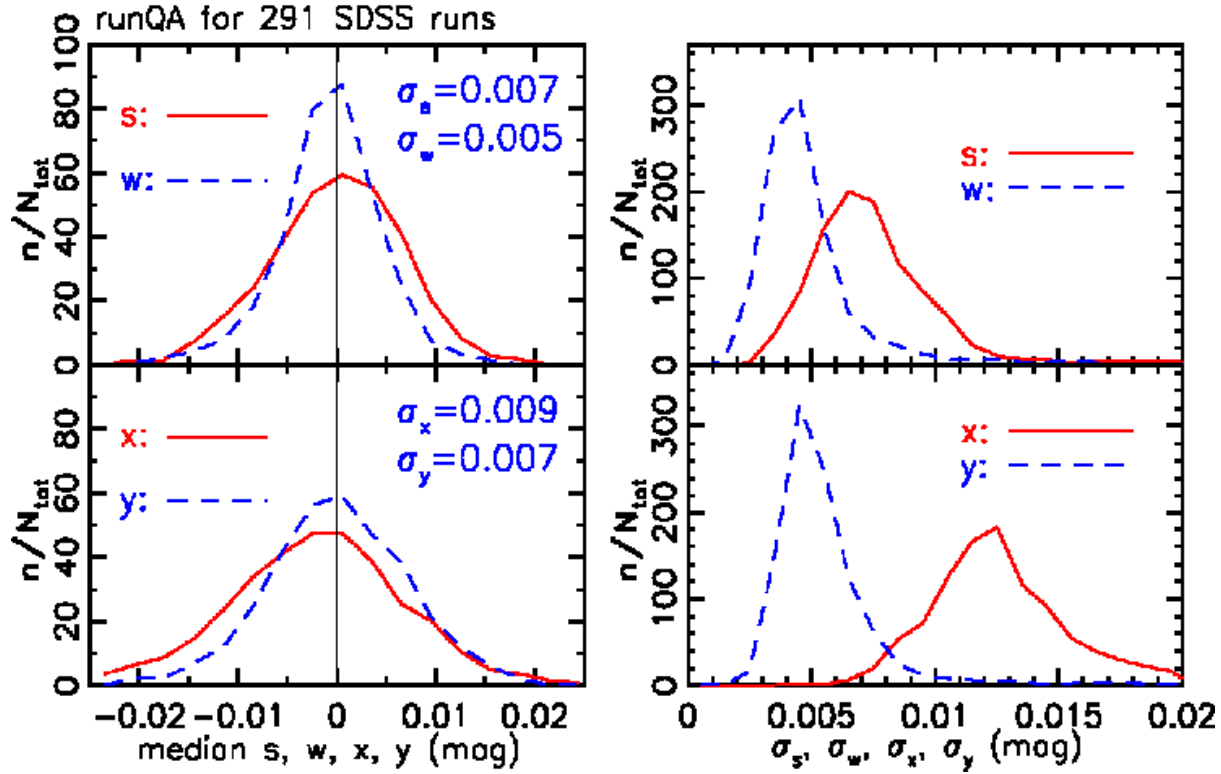}
   \caption{The two left panels show the distributions of the median for the 
    four principal colors (which measure the position of the stellar locus 
    position), evaluated for 291 SDSS runs processed to date ($\sigma$ values
    are the distribution rms widths).  The two right 
    panels show the distributions of the rms scatter of the mean stellar locus 
    position for six camera columns in a given run. The fact that the median values 
    of this rms scatter are similar to the widths of the distributions shown
    in the two left panels demonstrates that the observed variations in the locus 
    position are dominated by the errors in photometric zeropoint calibration. 
    } 
   \label{PCdistribution}
\end{figure*}

\subsubsection{ The distributions of SDSS photometric calibration errors }

\begin{figure}[t]
   \centering
   \includegraphics[angle=0,width=8cm]{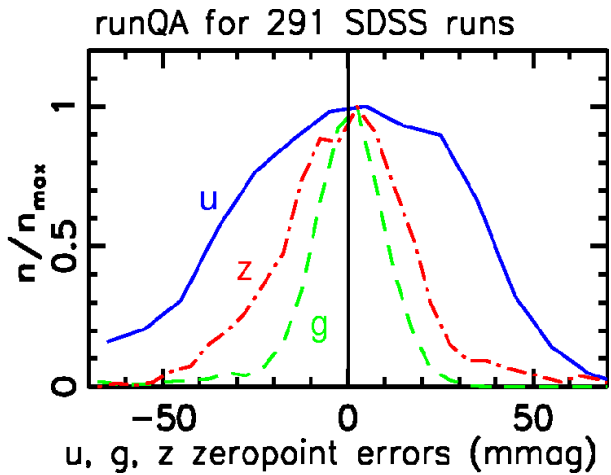}
   \caption{The distribution of the systematic errors in photometric 
    zeropoint for the $u$, $g$, and $z$ bands (the distributions for the $r$ and $i$ 
    bands are similar to the $g$ band distribution). The distribution widths
    are $\sim0.01$ mag in the $g$, $r$, and $i$ bands, 0.02 mag in the $z$
    band, and 0.03 mag in the $u$ band.
    } 
   \label{errors}
\end{figure}

The principal colors are insensitive to ``gray'' calibration errors (i.e. the
same offset in all bands). In order to estimate the photometric zeropoint
errors in each SDSS band, we use an ansatz motivated by the fact that 
these errors are typically the smallest in $g$, $r$ and $i$ bands, as indicated
by the direct comparison with calibration stars. With an assumption that 
the errors in the $g$, $r$ and $i$ bands sum to zero, we compute the photometric 
zeropoint errors in each SDSS band by inverting the equations that
define principal colors (by assuming that, instead, the $r$ band calibration
errors are zero, one gets practically the same solutions, in particular,
the $u$ band calibration error is about four times the offset in $s$ color). 

Fig.~\ref{errors} shows the distribution of inferred photometric 
zeropoint errors in three representative bands for 291 SDSS runs. 
The rms widths of these distributions are $\sim0.01$ mag in the $g$, $r$, 
and $i$ bands, 0.02 mag in the $z$ band, and 0.03 mag in the $u$ band. 
We emphasize that these values apply to all processed SDSS runs -- 
those with the worst calibration errors are excluded from the public
data releases (the rms widths remain similar to the above listed values).

\section{Astrometric calibration, star/galaxy classification, flatfield vectors, etc.}

Additional tasks performed by the {\it runQA} pipeline include an analysis of the accuracy 
of relative astrometric calibration\footnote{The accuracy of absolute and relative
astrometric calibration is discussed in detail by Pier et al. (2003).}
(transformations between positions measurements in the five SDSS bands), and analysis 
of median principal colors as 
a function of chip position to track the flatfield vector accuracy. For example, the 
latter was used to discover that (and to derive appropriate corrections for) 
the shapes of the flatfield vectors vary with time (on a scale of several dark 
runs), up to 20\% in some  $u$ band chips.

A similar pipeline, {\it matchQA}, compares two observations of the same area 
on the sky. It quantifies the repeatability of SDSS measurements, and provides
a handle on the sensitivity of measured parameters to varying observing 
conditions (such as seeing and sky brightness). For example, its
outputs demonstrate that the star/galaxy classification is typically repeatable 
at the $>99\%$ level at the bright end ($r<20$), and at the $>95\%$ level for sources
as faint as $r=21.5$. Furthermore, a direct comparison of photometry for 
multiply observed sources demonstrates that not only are the (random) photometric
errors small ($\sim$0.02 mag), but they themselves are accurately determined
by the photometric pipeline (see Fig. 2 in Ivezi\'{c} et al. 2003). Furthermore,
the tails of the photometric error distribution are well controlled and
practically Gaussian (see Fig. 3 in Ivezi\'{c} et al. 2003).

\section{   Discussion   }

SDSS is an excellent example of a modern astronomical survey -- it produces
unprecedentedly accurate data ($\sim$10 times more accurate than previous
large-scale optical sky surveys, such as POSS, see Sesar et al. 2004), 
with a large peak data rate (20 GB/hr), 
and is built upon sophisticated data acquisition, processing and distribution systems
($\sim$million lines of code), that were developed by a large number of collaborators 
($\sim$50). The success of SDSS provides encouragement that even more 
ambitious surveys, such as LSST (Tyson 2002), which is expected to produce
and immediately process 20 TB of data per observing night, may also be 
successful endeavors. 
 
SDSS also demonstrated, as described in this contribution, that a quantitative 
and efficient data quality assessment can be designed and implemented even when 
the ``true answers'' are not known for individual measurements. However, 
the method described here is not universal -- it applies only to the wavelength
range accessed by SDSS (0.3--1 $\mu$m), and to Galactic latitudes more
than $\sim$15 degree from the galactic plane. Despite these shortcomings, 
it is a robust automated method that tracks the accuracy of SDSS photometric
zeropoint calibration to better than 0.01 mag. We estimated, using this method,
that typical photometric zeropoint calibration errors for SDSS imaging
data are not larger than $\sim0.01$ mag in the $g$, $r$, and $i$ bands, 0.02 mag 
in the $z$ band, and 0.03 mag in the $u$ band.

\vskip 0.1in
\acknowledgements Funding for the creation and distribution of the SDSS Archive has been provided by the Alfred P. Sloan Foundation, the Participating Institutions, the National Aeronautics and Space Administration, the National Science Foundation, the U.S. Department of Energy, the Japanese Monbukagakusho, and the Max Planck Society. The SDSS Web site is http://www.sdss.org/.

The SDSS is managed by the Astrophysical Research Consortium (ARC) for the Participating Institutions. The Participating Institutions are The University of Chicago, Fermilab, the Institute for Advanced Study, the Japan Participation Group, The Johns Hopkins University, Los Alamos National Laboratory, the Max-Planck-Institute for Astronomy (MPIA), the Max-Planck-Institute for Astrophysics (MPA), New Mexico State University, University of Pittsburgh, Princeton University, the United States Naval Observatory, and the University of Washington.

\v{Z}I thanks Princeton University for generous 
financial support.


\end{document}